# $V_N C_B$ defect as source of single photon emission from hexagonal boron nitride

A. Sajid[1,2,3,*] and Kristian S. Thygesen[1]

[1]CAMD, Department of Physics, Technical University of Denmark, 2800 Kgs. Lyngby, Denmark.

[2]Department of Physics, GC University Faisalabad, Allama Iqbal Road, 38000 Faisalabad, Pakistan.

[3]University of Technology Sydney, School of Mathematical and Physical Sciences, Ultimo, New South Wales 2007, Australia.

*Corresponding Author:  sajal@dtu.dk

**Abstract:** Single photon emitters in 2D hexagonal boron nitride (hBN) have attracted a considerable attention because of their highly intense, stable, and strain-tunable emission. However, the precise source of this emission, in particular the detailed atomistic structure of the involved crystal defect, remains unknown. In this work, we present first-principles calculations of the vibrationally resolved optical fingerprint of the spin-triplet (2) $^3B_1$ to (1) $^3B_1$ transition of the $V_N C_B$ point defect in hBN. Based on the excellent agreement with experiments for key spectroscopic quantities such as the emission frequency and polarization, the photoluminescence (PL) line shape, Huang-Rhys factor, Debye-Waller factor, and re-organization energy, we conclusively assign the observed single photon emission at ~2eV to the $V_N C_B$ defect. Our work thereby resolves a long-standing debate about the exact chemical nature of the source of single photon emission from hBN and establishes the microscopic understanding necessary for controlling and applying such photons for quantum technological applications.

Keywords:  Single photon emission, defects in h-BN, PL line shape, Huang Rhys factor

## Introduction:

The discovery [1-4] of single-photon emission (SPE) from defects in hexagonal boron nitride (hBN) has stimulated significant experimental and theoretical research in an effort to identify the chemical origin of this emission. Much of the motivation stems from the possibility of using such SPEs as the basis for novel nanophotonic- or quantum information technologies [5-8]. Their exploitation, however, depends on a thorough understanding of the chemical nature of the colour centres in hBN and the ability to control and tune their structural arrangements and spectroscopic properties. While various colour centres have been proposed as potential sources of SPEs in hBN [9-12], only a few defects have been conclusively identified [10,13,14]. Many attempts have been made to predict the exact nature of the defect causing emission in the visible part of the spectrum, i.e. in the vicinity of 2 eV [4,9-11,13,15,16]. However, no conclusive evidence for the nature of the defect in the form of theoretically predicted properties matching the experimental data, has so far been presented [10].

The emission frequencies of SPEs observed in hBN can be spread over a considerable spectral range from 1.4 to 2.2 eV. The spectra overall consist of a relatively sharp spectral line separated from a smaller broader feature, typically 150–200 meV lower in energy. These features are usually assumed to be the zero phonon line (ZPL) and associated phonon side bands (PSB), respectively, originating from a localized emission source, most likely involving at least one defect level lying deep within the semiconductor band gap. The PL spectra for a particular type of emitter is signature of specific chemical structure of the defect. Therefore, one can identify the nature of an emitting center by calculating its PL spectra and comparing it to the experiment.

In an initial survey study employing density functional theory (DFT) with a gradient corrected (GGA) functional, a carbon based defect $V_NC_B$ was proposed as source of the observed emission. This assignment was based upon a comparison of calculated spectral quantities, e.g. the Huang Rhys (HR) factor, photoluminescence (PL) line shape and zero phonon line (ZPL) energies [4], with experimental data. However, the hBN community could not conclusively agree on this prediction and the source of the observed emission remained under considerable debate[10]. The dissensus was partly due to a lack of experimental studies of controlled incorporation of carbon impurities in hBN. Another reason was the, after all, rather poor agreement between theoretical calculations and experiment (both quantitatively e.g. ZPL energies and qualitatively e.g. PL line shape) [10].

In a recent experimental study, the observation of an increased density of emitters with ZPL close to 2 eV following controlled incorporation of carbon into h-BN is reported [17]. The work tentatively assigns the emission to a $V_NC_B$ defect based upon previous theoretical reports [4,10,13,15,16,18] none of which, however, provide key spectroscopic observables in agreement with the new experiments.

In this work, we present conclusive evidence that the photons emitted in the 2 eV spectral range and reported in Ref. [17] are due to transitions between triplet states of the $V_NC_B$ point defect. Our first-principles calculations employing the HSE06 range separated hybrid functional show an excellent agreement with the recent experimental results for the luminescence line shape, Huang-Rhys factor, Debye-Waller factor and the reorganization energy. We emphasize that despite of numerous previous theoretical investigations of the $V_NC_B$ defect, this is the first time that convincing agreement between theory and experiment is obtained for the 2 eV SPE in hBN.

**Theory and Results:**

The intensity $I(h\nu)$ of a photo-luminescence (PL) emission spectrum under the Frank-Condon approximation (i.e. the transition dipole moment ($\hat{\mathbf{d}}$) is similar in the initial and final configurations), are given by the following expression

$$I(h\nu) = \frac{64\pi^4 \nu^3}{3hc^3} [\hat{\mathbf{d}}]^2 |\langle \Psi_i | \Psi_f \rangle|^2 \, \delta(E_f - E_i - h\nu). \qquad (1)$$

Here $\Psi_i$ and $\Psi_f$ are the vibrational states of the initial and final configurations involved in the emission process (localized defect states in our case) as shown in Fig 1.

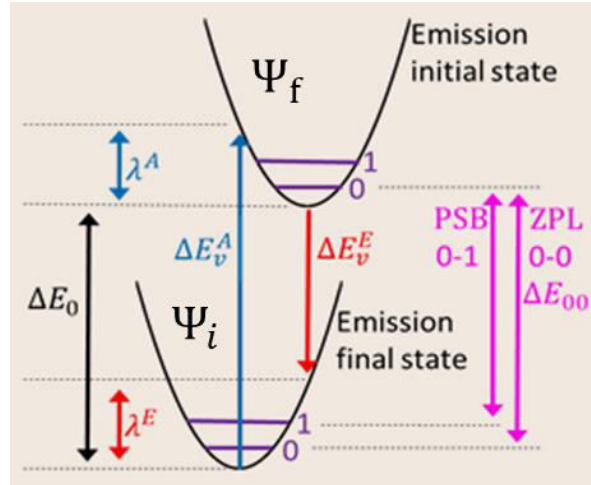

**Fig. 1.** One-dimensional configuration coordinate diagram depicting different quantities used to describe the process of light emission from defect states.

Fig. 1 illustrates the process of light absorption and emission due to transitions between two localized electronic states. Incoming photons with an average energy, $\Delta E_v^A$, produce the excited state. Excess energy equal to the absorption reorganization energy $\lambda^A$ is then dissipated as heat through energy transfer into the vibrational modes of the hBN lattice followed by the emission process of interest. This emission occurs at average energy $\Delta E_v^E$, followed by heat loss equal to the emission reorganization energy $\lambda^E$ and return to the original equilibrium state. The sum of the two reorganization energies $\lambda^A + \lambda^E$ equals the energy difference between the average absorption and emission energies $\Delta E_v^A - \Delta E_v^E$ and is known as the *Stokes shift*, with the average transition energies $\Delta E_v^A$ and $\Delta E_v^E$ usually referred to as *vertical* transition energies. The latter are readily calculated using DFT by evaluating the energies of both states at the two equilibrium geometries and can be compared directly with values extracted from photoluminescence spectra. The difference in energy between the two equilibrium geometries, i.e. the vertical distance between the minima of the two parabolas in Figure 1, is called the *adiabatic* transition energy $\Delta E_0$. When corrected for the zero-point motion of each state, this gives the energy of the ZPL, $\Delta E_{00}$. This zero phonon line along with the phonon side bands associated with emission or absorption of a phonon, are the quantities measured experimentally in PL or resonance excitation spectra. The PL spectrum thus provides a unique signature of a defect and, in principle, comparison of the calculated spectrum with the experimentally measured one can help to identify the source of emission.

In order to calculate the luminescence from Eq. (1) the vibrational overlap integrals $\langle \Psi_i | \Psi_f \rangle$, known as Frank-Condon Integrals, must be evaluated. In general, this poses a significant challenge due to the large number of atoms involved in the vibrations that can couple to the electronic transitions, and the fact that normal modes in the ground- and excited states are generally different [19,20]. To overcome the problem we apply the linear transformation approach of Duschinsky [19], in which the normal modes in the initial state are related to the normal modes in the final state via the affine transformation,

$$Q_i = JQ_f + K_{i,f} \qquad (2)$$

In this expression, the Duschinsky matrix $J = (L_i)^{-1} L_f$, is composed of the transformation matrices from normal coordinates to mass-weighted Cartesian coordinates and the vector $K_{i,f}$ contain the displacement between the initial and final states expressed in the basis of initial state normal modes. It is given by $K_{i,f} = (L_{i,f})^{-1} M^{\frac{1}{2}} \Delta X_{i,f}$, where M is diagonal matrix of atomic masses and $\Delta X_{i,f}$ is a vector representing the shift of nuclear Cartesian coordinates between the initial and final states. By utilizing the Duschinsky transformation, the Frank-Condon integral $\langle \Psi_i | \Psi_f \rangle$ can be determined using the well-known expression for the wave function overlap of displaced 1D harmonic oscillators. In practice, the displacements $K_{i,f}$ are expressed in units of the zero-point lengths of each mode. A displacement of 1 in any mode raises the re-organization energy $\lambda_j$ (subscript j represents the mode number) by $h\upsilon_j/2$ according to the following equation.

$$\lambda_j = \frac{K_{i,f}^2}{2} h\upsilon_j \qquad (3)$$

The total re-organization energy $\lambda$ and Huang-Rhys factor $S^{E,A}$ ($S^E$ for HR factor during emission and $S^A$ for absorption) are then given by

$$\lambda = \sum_{j=1}^{3n-3} \lambda_j \quad , \quad S^{E,A} = \sum_{j=1}^{3n-3} \frac{K_{i,f}^2}{2} \qquad (4)$$

We have used the theory outlined above to calculate the PL line shape for the $V_N C_B$ defect. Before presenting our results, we briefly discuss the basic atomic and electronic structure of $V_N C_B$.

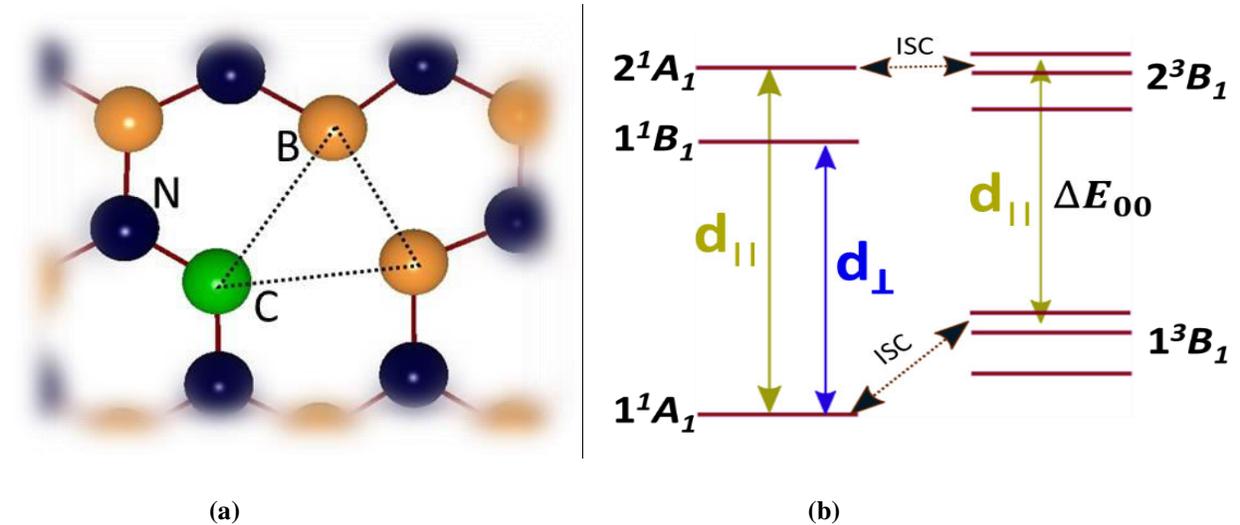

(a)          (b)

**Fig. 2.** (a) Defect $V_N C_B$ with $C_{2v}$ symmetry embedded in a periodic h-BN sheet. (b) Schematic energy level structure of $V_N C_B$. This defect has a singlet ground state, with the emission observed around ~2eV here assigned to the (2) $^3B_1$ to (1) $^3B_1$ transition.

Fig. 2(a) shows the atomic structure of the $V_NC_B$ point defect in an hBN sheet. This defect has previously been studied in detail and many of its properties, e.g. symmetry, dipole allowed transitions, and allowed inter-system crossing due to spin-orbit coupling, have been discussed [16]. However, conclusive evidence for this defect being the source of SPE in the visible frequency range is still missing. The defect has $C_{2v}$ symmetry and a singlet (1) $^1A_1$ ground state. After excitation from the ground state to the phonon sideband of the optically bright (2) $^1A_1$ state, the system can relax back to the (1) $^1A_1$ ground state by radiative emission of an in-plane polarized photon or vibronically to the (1) $^1B_1$ state and then back to the (1) $^1A_1$ via emission of an out-of-plane polarized photon. Alternatively, inter-system crossing enabled by the spin-orbit coupling can take place resulting in decay of (2) $^1A_1$ population to (2) $^3B_1$ [10,16]. The population of (2) $^3B_1$ can then decay to (1) $^3B_1$. It is the transition from (2) $^3B_1$ to (1) $^3B_1$ that we propose is responsible for the observed emission around ~2eV. The transition into the triplet channel depends on the competition between the radiative transitions within the singlet manifold i.e. (2) $^1A_1$ to (1) $^1A_1$ and the inter-system crossing, respectively. Detailed calculations of these rates will be the topic of future work.

To quantify the effect of electron-phonon coupling on the (2) $^3B_1$ to (1) $^3B_1$ transition, we calculate the Huang-Rhys factor $S^E$(Eq. 4), which is a measure of the average number of phonons emitted during an optical transition. The Huang-Rhys factor is related to the Debye-Waller factor $\alpha^E$ (the fraction of light emitted into the ZPL) via $\alpha^E \approx e^{-S^E}$. Our calculated values for the (2) $^3B_1$ to (1) $^3B_1$ transition for the adiabatic energy ($\Delta E_0$), Huang Rhys factor($S^E$), Debye-Waller factor ($\alpha^E$) and re-organization energy($\lambda^E$) are shown in Table 1, along with the corresponding values extracted from the experimental spectra in Fig. 1(d) of Ref. [17]. A detailed description of the procedure used to extract the experimental values is provided in the supporting information (S.I.) section S. 1.

**Table 1.** Comparison of calculated values of transition energy ($\Delta E_0$), Huang Rhys factor ($S^E$), Debye Waller factor ($\alpha^E$) and re-organization energy($\lambda^E$) for $V_NC_B$ defect compared with the values extracted form experimental spectra [17].

|  | $\Delta E_0$ / eV | $S^E$ | $\alpha^E$ | $\lambda^E$ / eV |
|---|---|---|---|---|
| This work | 1.95 | 1.50 | 0.21 | 0.11 |
| Experiment[17] | 2.10 | 1.45 | 0.23 | 0.11 |

Table 1 reveals an excellent agreement between calculations and experiment, essentially confirming that the emission observed in the experiment is associated with the $(2)\,^3B_1$ to $(1)\,^3B_1$ transition of the defect $V_NC_B$. The difference between our calculated values and previous works [4] may be assigned to the use of different xc-functionals (HSE06 vs. PBE) and numerical parameters (higher energy cut offs and more stringent convergence criteria are used in this work). Our calculated luminescence line shape for the $2^3B_1$ to $1^3B_1$ transition is shown in Fig. 3 together with the experimental PL spectrum from Ref. [17]. Taking into account that no fitting parameters, other than a spectral broadening of 0.02eV and a rigid shift of 0.15 eV to match the ZPL peak, have been used to obtain the luminescence spectrum, the excellent agreement with experiment further supports that this emission is indeed produced by the $(2)\,^3B_1$ to $(1)\,^3B_1$ transition of the $V_NC_B$ defect.

Since the defect $V_NC_B$ maintains its $C_{2v}$ symmetry in both the triplet $(1)\,^3B_1$ ground state and the $(2)\,^3B_1$ excited state, only fully symmetric $a_1$ modes ( i.e. the modes that do not distort the $C_{2v}$ symmetry of the defect) contribute to the luminescence spectrum because only these modes yield non-zero Franck-Condon integrals in Eq. (1). The dominant contribution to the phonon side bands (PSB) stems from the defect breathing modes. Since the modes contributing to the PSB change the electronic polarizability of the system, these are Raman-active modes. We note in passing that it would be interesting to explore whether effects from C-based defects are visible in the h-BN Raman spectrum.

Overall, the phonon modes that contribute to photoluminescence are not very localized at the $V_NC_B$ defect and there is a major contribution from bulk-like modes to the PSB (Fig S.1 shows some modes with large contribution to the PSB). This is an interesting feature, and further studies should be directed towards understanding the physics of electron-phonon coupling in this material, to be able to utilize these emitters in practical applications.

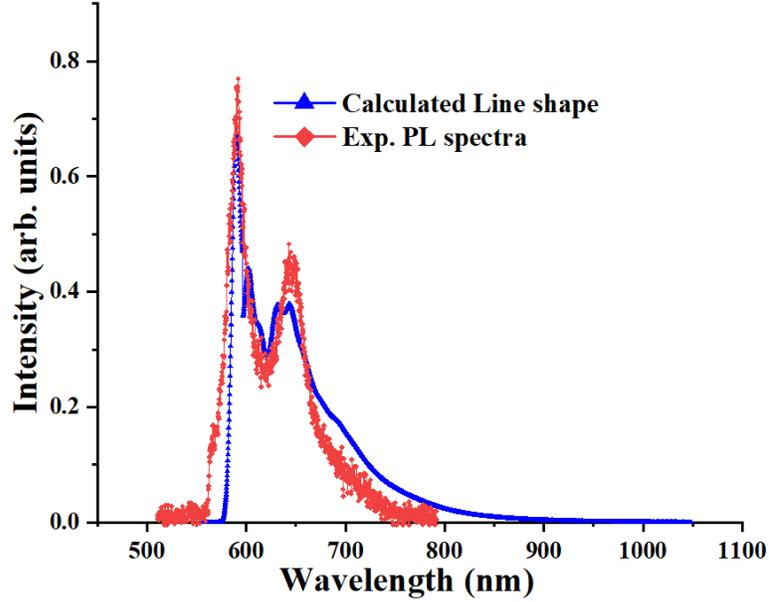

**Fig. 3.** Calculated Photoluminescence spectrum fitted to experimental spectrum[17]. For comparison, calculated spectra with adiabatic ZPL energy $\Delta E_0 = 1.95 eV$ was aligned to the position of ZPL at 590 nm (2.10eV) was aligned to match to the highest intensities of the experimental spectrum which corresponds to about 0.15 eV blueshift compared to the calculated value. This difference is within the usual inaccuracy of the DFT method.

We have also calculated the absorption spectrum of the triplet transition in the $V_N C_B$ defect, see the S.I. Section S. 4. (We note that this absorption spectrum should not be compared to the experimental absorption spectrum of the defect, which is determined by transitions between the singlet states, see Fig. 2(b). It is, however, instructive to consider the absorption between the triplet states). An important feature to note is that the calculated absorption spectrum is not a perfect mirror image of the emission spectrum. The cause of some asymmetry between absorption and emission spectrum is associated with the Duschinsky matrices. Duschinsky matrices $J$, for $(2)\ ^3B_1 \rightleftarrows (1)\ ^3B_1$ are not unitary matrices. In other words, one cannot make the simple assumption that modes in the ground and excited state are similar for the case of $(2)\ ^3B_1 \rightleftarrows (1)\ ^3B_1$ transition. This assumption has previous been made for calculation of the emission spectrum of this transition, and hence a poor agreement with the experiment was seen [4].

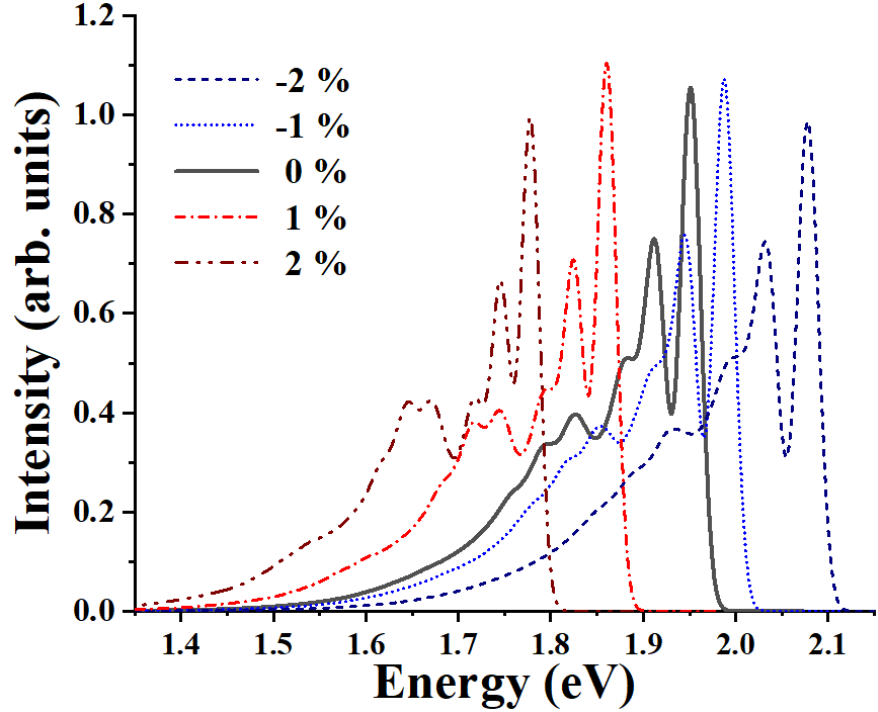

**Fig. 4.** Calculated PL spectrum of the defect $V_NC_B$ for (2) $^3B_1$ to (1) $^3B_1$ transition as a functional of compressive (negative) and tensile (positive) strain.

Since strain tunability is a major feature of quantum emitters in h-BN [10,18], we simulate the vibrationally resolved PL spectra for the (2) $^3B_1$ to (1) $^3B_1$ transition in $V_NC_B$ at different values of compressive and tensile strain, see Fig. 4. One can note that the strain not only modifies the ZPL energy, but also changes the shape of PSB (although only slightly, larger strain can, however, cause a large change) and hence the Huang-Rhys factors. We thus predict that along with the ZPL energies, the Huang-Rhys factors of these emitters can also be changed with applied strain and future experiments should be directed to explore this feature. Our calculated values of Huang-Rhys factors as a function of applied strain are provided in S.I table S. 6.

Finally, a word on why we exclude other carbon related defect species as the possible source of the observed SPE. Defect species such as $C_B$ and $C_N$ in both neutral and charged states do not feature energy levels in the band gap with the right energy separation to produce an emission around ~2eV [12]. Similarly, the defect $V_BC_N$ can be excluded as its adiabatic transition has dipole orientation ($d_\perp$) perpendicular to the h-BN plane, therefore should emit light in the plane of h-BN sheet in contrast to the experimental findings. We have also investigated

Stone-Wall type defects involving a carbon atom namely $S_WC_B$ and $S_WC_N$ [20]. Only $S_WC_B$ can produce an in-plane polarized transition in the right energy window. However, its re-organization energy is too large to match the observed narrow PL line shape and relatively large Debye-Waller factor. Specifically, our calculated adiabatic energy for $S_WC_B$ is 1.9eV while we obtain a re-organization energy for this transition of 0.4eV making it unlikely to be the source of the observed emission. We also considered the positive and negative charge states of the defect $V_NC_B$ and found that only $V_NC_B^{+1}$ has a transition $((1)\,^2B_2$ to $(1)\,^2A_1)$ in the desired energy window, but re-organization energy for this transition is too high to match the experiment. A table of re-organization energies and adiabatic transition energies of the most relevant transitions in a select set of defect candidate systems is presented in the S.I. section S. 5. From this overview, it follows that among all the carbon related defects, only the defect $V_NC_B$ in the triplet channel present an adiabatic energy and re-organization energy consistent with the experimental data, and must therefore be considered to be the most likely source of the observed emission.

**Conclusion**: Based on an excellent agreement between our first-principles calculations and recent experimental photoluminescence data for carbon-containing defects in hBN, we have identified the detailed atomic structure of the color center responsible for the highly intense 2 eV photons emitted from hBN. Our calculated transition energy, Huang-Rhys factor Debye-Waller factor and re-organization energy for the triplet transition of the defect $V_NC_B$ are all found to be in very good quantitative agreement with the experimental data. In addition, the calculated PL lineshape for this transition shows a high similarity with the measured spectrum. We have traced the origin of the high intensity and narrow line shape of the $V_NC_B$ emitter to the delocalized nature of the vibrational modes coupling to the electronic transition, which result in a strongly reduced reorganization energy. Our results resolve a long-standing debate about the chemical nature of these color centers, and establishes the basic microscopic understanding of the photo-physical properties of these emitters thereby paving the way to their future deployment for e.g. quantum technological applications.

**Computational parameters:**

Calculations are performed for periodically replicated defects in 2-D h-BN monolayer. For calculation of total energy, electronic structure and ground state geometry we used version 5.3.3 of the Vienna Ab Initio Simulation Package (VASP)[21,22]. For accurate calculation of electron spin density close to the nuclei, the projector augmented wave method (PAW) [23,24] was applied together with a plane wave basis set. We utilized the standard PAW-projectors provided by the VASP package. Pristine single-layer h-BN was first geometrically optimized using the conventional cell and a 27×27×1 Monkhurst-Pack reciprocal space grid. A large vacuum region of 30 Å width was used to separate a single layer of h-BN from its periodic images and to ensure that interaction between periodic images is negligible. The optimized bond length of pristine h-BN is 1.452 Å. All the defects were then realized in a 9x9x1 supercell and allowed to fully relax using a plane wave cut-off of 700 eV for a maximum force of 0.001 eVÅ$^{-1}$. K-point convergence was check and finally a k-point mesh of 3x3x1 was used for all calculations. The normal modes and dynamical matrices were calculated at Gamma point of the BZ. The total energies of the excited states were calculated within the ΔSCF method that provides accurate zero-phonon-line (ZPL) energy and Stokes-shift for the optical excitation spectra for triplet manifold [16]. The PL line shape was calculated using Duschinsky linear transformation method [19,25]. Further details of our calculation of vibrationally resolved PL spectra are provided in the S.I. section S. 2.

## Acknowledgment

This work was supported by the Center for Nanostructured Graphene (CNG) under the Danish National Research Foundation (project DNRF103) and the European Research Council (ERC) under the European Union's Horizon 2020 research and innovation program (Grant No. 773122, LIMA). Authors thank Noah Mendelson for providing experimental PL data for comparison.